\documentclass[11pt,a4paper]{article} 
\usepackage{authblk}
\usepackage{graphicx}%
\usepackage{amsmath,amssymb,amsfonts}%
\usepackage{amsthm}%
\usepackage{mathrsfs}%
\usepackage[title]{appendix}%
\usepackage{xcolor}%
\usepackage{textcomp}%
\usepackage{manyfoot}%
\usepackage{booktabs}
\usepackage{algorithm}%
\usepackage{algorithmicx}%
\usepackage{algpseudocode}%
\usepackage{listings}%
\usepackage{algorithm}
\usepackage{array}
\usepackage{textcomp}
\usepackage{stfloats}
\usepackage{verbatim}
\usepackage{stackrel}
\usepackage{subcaption}
\usepackage{epsf}
\usepackage{color}
\usepackage{multirow,tabularx}
\usepackage{ifthen}
\usepackage{tikz}
\usetikzlibrary{shapes.geometric}
\usepackage{pgf-pie}
\usepackage{times}

\usepackage{amssymb,amsmath,amsthm}

\usepackage{xcolor}
\usepackage{fancyhdr}    
\usepackage{etoolbox}    

\usepackage[hidelinks]{hyperref} 

\usepackage{textcomp}    
\usepackage{microtype}   
\usepackage{lipsum}      

{\normalfont\huge\bfseries\centering}

\newcommand{\define}{\triangleq}

\newcommand{\beq}{\begin{equation}}
	\newcommand{\eeq}{\end{equation}}

\newcommand{\expect}[1]{{\bf E}\left[{#1}\right]}

\newcommand{\var}[1]{{\bf var}\left[{#1}\right]}
\newcommand{\brac}[1]{\left({#1}\right)}

\hypersetup{
	colorlinks=false,
	pdfauthor={Prathamesh Anwekar, Kaushal Kirpekar, Mahesh B, Sainath Bitragunta},
	pdftitle={Probabilistic Analysis of Various Squash Shots and Skill Study of Different Levels of Squash Players and Teams},
	pdfkeywords={squash, sports analytics, probabilistic modeling}
}

\begin{document}
	
\title{Probabilistic Analysis of Various Squash Shots and Skill Study of Different Levels of Squash Players and Teams}

\author[1]{Prathamesh Anwekar\thanks{Formerly with the Department of Electrical and Electronics Engineering, Birla Institute of Technology and Science Pilani, Vidya Vihar, Pilani 333031, Rajasthan, India.}}
\author[1]{Kaushal Kirpekar\thanks{Formerly with the Department of Electrical and Electronics Engineering, Birla Institute of Technology and Science Pilani, Vidya Vihar, Pilani 333031, Rajasthan, India.}}
\author[2]{Mahesh B\thanks{Department of Mathematics, Visvodaya Technical Academy, Udayagiri Road, Kavali 524201, Andhra Pradesh, India.}}
\author[1]{Sainath Bitragunta\thanks{Department of Electrical and Electronics Engineering, Birla Institute of Technology and Science Pilani, Vidya Vihar, Pilani 333031, Rajasthan, India. Email: \texttt{sainath.bitragunta@pilani.bits-pilani.ac.in, lsnb290580@gmail.com. Corresponding author.}}}

\affil[1]{Department of Electrical and Electronics Engineering, BITS Pilani, Rajasthan, India}
\affil[2]{Department of Mathematics, Visvodaya Technical Academy, Kavali, Andhra Pradesh, India}

\date{} 
	
	\maketitle

\begin{abstract}
	We introduce a compact probabilistic model for two-player and two-team (four-player) squash matches, along with a practical skill-comparison rule derived from point-scoring probabilities. Using recorded shot types and court locations, we analyze how shot distributions differ between professional-level and intermediate-level players. Our analysis shows that professional players use a wider variety of shots and favor backcourt play to maintain control, while intermediate players concentrate more on mid-court shots, generate more errors, and exercise less positional control. These results quantify strategic differences in squash, offer a simple method to compare player and team skill, and provide actionable insights for sports analytics and coaching.
\end{abstract}


\section{Introduction}
\label{sec:intro}

Regular participation in sports reduces stress, improves fitness, and fosters competitiveness. Racket sports such as tennis, badminton, and squash combine cardiovascular exercise with rapid decision-making and fine motor control. Squash, in particular, is a high-intensity court sport played by millions worldwide on courts bounded by walls. Due to its physical and tactical demands, squash lends itself well to quantitative analysis.

\begin{figure}[htb!]
	\centering
	\includegraphics[width=1\linewidth]{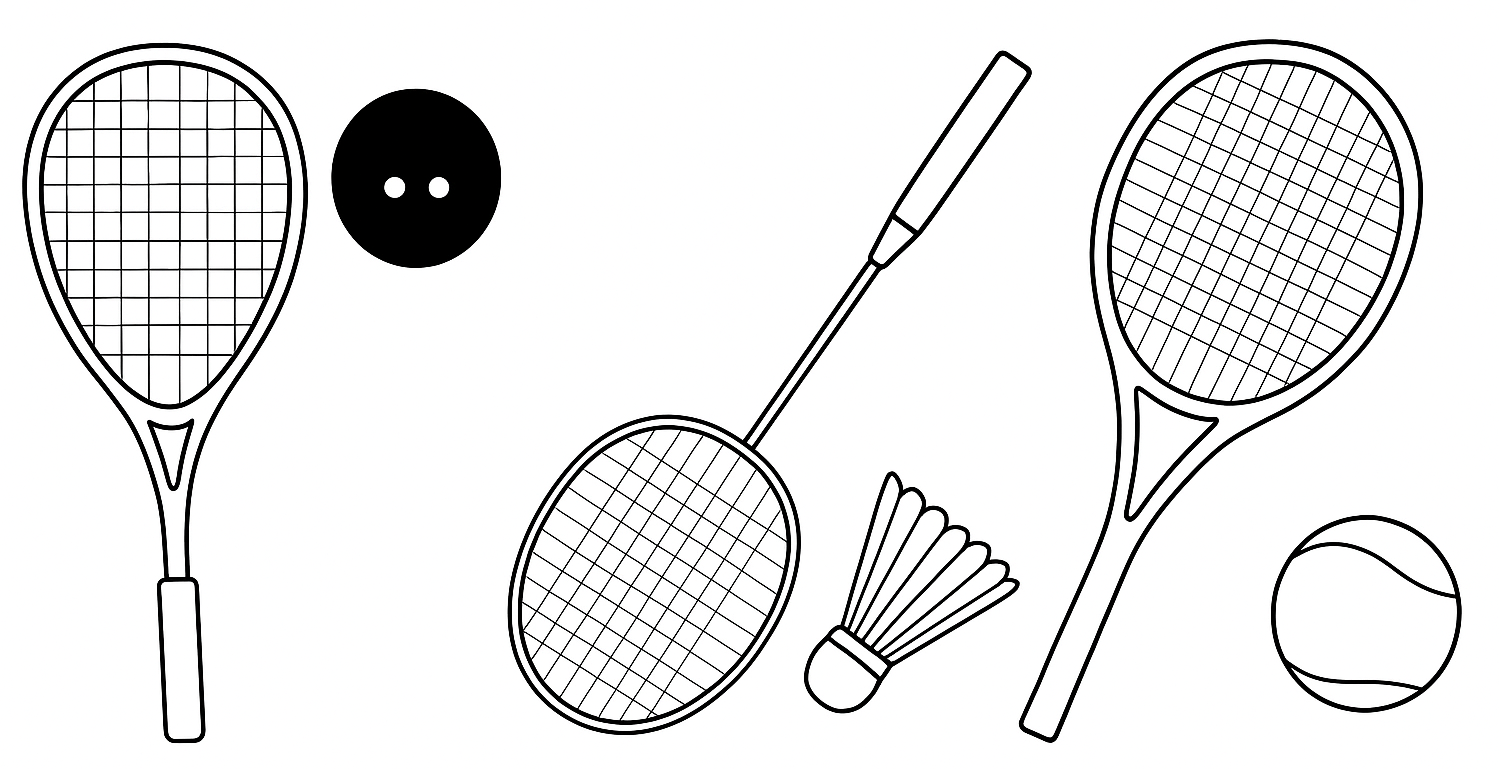}
	\caption{Illustration of three popular racket sports and their main equipment. A squash racket and ball are shown on the left, a badminton racket and shuttlecock in the center, and a tennis racket and ball on the right. Court structure and rules differ significantly across these sports.}
	\label{fig:racket_sports_illu}
\end{figure}

Recent advances in artificial intelligence, computer vision, and data science enable the detailed analysis of sports and probabilistic modeling of play. Despite these advances, data-driven probabilistic analyses of squash remain limited. In this paper, we introduce a compact probabilistic model for two-player and two-team (four-player) squash, and we propose a simple skill-comparison rule based on point-scoring probabilities. We also investigate how shot types are distributed across court regions and how these distributions differ between professional and intermediate players.

We collected and structured a dataset of real matches, recording the types of shots and the locations of the courts. Using statistical and computational tools, we analyze shot distributions, compare strategies across skill levels, and derive the distribution for scoring $k$ points given a player's single score point winning probability. Our analysis reveals that professional players employ a broader range of shots and favor backcourt play to maintain control over the game. In contrast, intermediate players focus more on mid-court shots, commit more errors, and exhibit weaker positional control. These findings quantify strategic differences in squash and provide actionable insights for coaches and sports analysts.

\paragraph{Methodology.} We follow a three-step methodology: (i) collect and structure play data; (ii) perform statistical analysis; and (iii) interpret results for coaching and analytics.

\paragraph{Scope and assumptions.} To keep the analysis tractable, we make the following assumptions:
\begin{enumerate}
	\item Lets and strokes are excluded from the analysis.
	\item Service shots are excluded.
	\item Volley shots are not treated as a separate category.
	\item All players in the dataset are right-handed.
	\item Shots of the same type from the same court region are treated as identical events.
\end{enumerate}

{\em Organization:} The remainder of this paper is organized as follows. Section~\ref{sec:court_n_shots} provides a brief overview of court layout, game description, and different squash shots.  Section~\ref{sec:prob_anal} presents the shot event modeling framework and analysis. Section~\ref{sec:sim_res} describes the case study analysis of shot type. Section~\ref{sec:Analysis_output} presents the quantitative analysis of shot output. Section~\ref{sec:conclusions} concludes and outlines future work.

\section{Understanding the Court and Different Shots}
\label{sec:court_n_shots}

To structure our analysis, we partition the squash court into regions and classify shots by technique and outcome. This section defines the court regions used in our dataset, summarizes the two-player game flow relevant to our data collection, and lists the shot types and outcome categories.

\subsection{Court regions and dimensions}

\begin{figure}[htb]
	\centering
	\includegraphics[width=\linewidth]{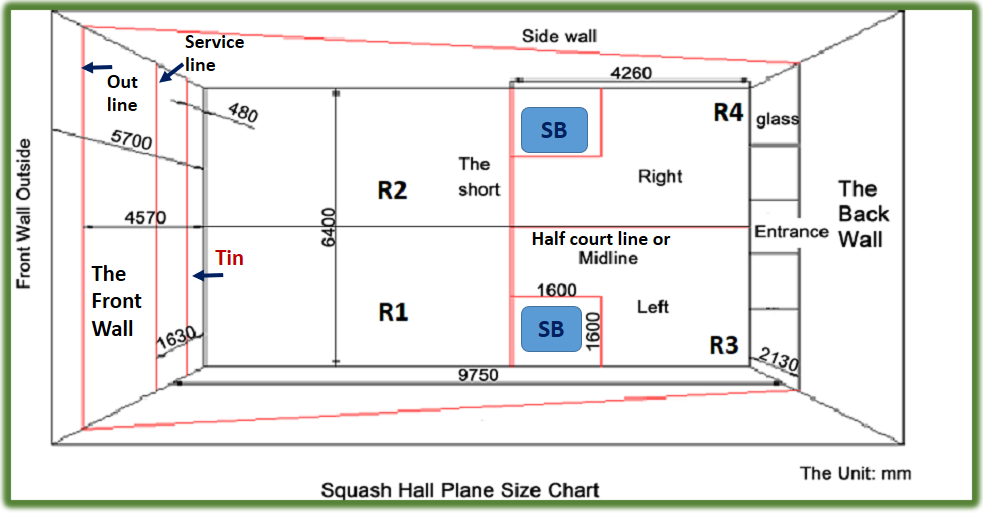}
	\caption{Court dimensions and region partitioning used in this study. SB: service box; R1--R4: court regions for shot-location analysis.}
	\label{fig:squash_court}
\end{figure}

Figure~\ref{fig:squash_court} shows the standard squash court layout and the four regions we use in this work. For reference, we use a court of length $9.75\,$m and width $6.4\,$m; the front-wall heights and service markings follow standard specifications~\cite{squash_court_dimensions,squash_court_diagram}. For analysis, we divide the floor plan into four regions (R1--R4), as indicated in the figure. These regions provide the spatial bins used when counting and comparing shot distributions.

\subsection{Two-player game flow (summary)}

For reproducibility, we recorded only rally shots. A concise summary of the two-player flow relevant to our dataset:

\begin{enumerate}
	\item One player serves to start a rally.
	\item The receiver returns the serve after zero or one bounce; thereafter, players alternate hitting the ball.
	\item Play continues until the rally ends (e.g., ball bounces twice, is out of bounds, or a let/stroke is called).
	\item We record rally events and the player who won the rally. 
\end{enumerate}

{\em Remark:} lets, strokes, and service actions are treated according to the assumptions listed in Section~\ref{sec:intro} and are excluded from the shot-distribution counts.

\subsection{Shot taxonomy and outcome categories}

We classify shots by hand (forehand/backhand), by technique, and by outcome. The 12 shot types in our taxonomy are:

\medskip
\noindent\textbf{Forehand shots (FHS):} FH parallel drive, FH cross drive, FH lob, FH boast, FH back wall, FH drop.\\
\textbf{Backhand shots (BHS):} BH parallel drive, BH cross drive, BH lob, BH boast, BH back wall, BH drop.

\medskip
\noindent Each recorded shot receives one outcome label from the following set:
\begin{itemize}
	\item \textbf{Normal (N)}: rally continues after the shot;
	\item \textbf{Winner (W)}: the shot ends the rally and wins the point directly;
	\item \textbf{Forced error (F)}: the opponent's error is provoked by the shot (opponent under pressure);
	\item \textbf{Unforced error (U)}: the player makes an error that is not attributable to opponent pressure.
\end{itemize}

\noindent In the next section, we present a probabilistic model of shot events and provide an insightful analysis. Furthermore, we also present modeling and skill Comparison of teams. To obtain more insights, we also present a numerical result plot. 

In later sections, we use the shot-type, region (R1--R4), and outcome labels to compute region-wise shot distributions and to compare strategy and error patterns between professional-level and intermediate-level players. 

\section{Squash shot event modeling and analysis}
\label{sec:prob_anal}

We consider the two-player squash game. Let $S_{1}$ denote the number of squash shots played by player$-1$. We define an indicator random variable as follows: If player$-1$'s squash shot is successful, that is, scores a point, $I_{s1} = 1$. Let this probability be $p_1$. Else $I_{s1} = 0$ with probability $1-p_{1}$. Note that $0 < p_{1} < 1$ and $0 < p_{2} < 1$.

Let $W_1 \define \sum_{s1 = 0}^{S_1} I_{s1}$. We see that $W_1$ follows a binomial distribution. We have $\expect{W_1} = S_{1}p_{1}$, $\var{W_1} =  S_{1}p_{1} (1-p_{1})$. We are interested in determining the probability of scoring $k$ points. Mathematically, we write $	\mathrm{P}(W_1 > k)$. Assuming Gaussian approximation~\cite{Papoulis_book} for large $S_{1}$, we have 
\begin{equation}
	\label{eq:kPt_score_prbty_P1}
	\mathrm{P}(W_1 > k) \approx Q\brac{\frac{k - S_{1}p_{1}}{\sqrt{S_{1}p_{1} (1-p_{1})}}},
\end{equation}
where $Q(.)$ denotes the Q$-$function defined on Normal distribution.

Similarly, for player$-2$, we can show that
\begin{equation}
	\label{eq:kPt_score_prbty_P2}
	\mathrm{P}(W_{2} > k) \approx Q\brac{\frac{k - S_{2}p_{2}}{\sqrt{S_{2}p_{2} (1-p_{2})}}}.
\end{equation}

\begin{figure}[h!]
	\centering
	\includegraphics[width=1\linewidth]{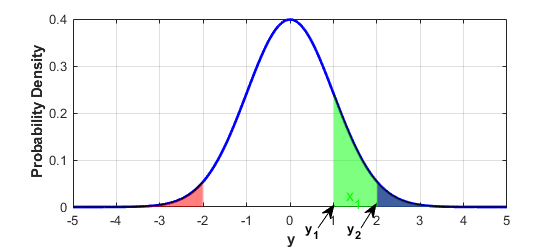}
	\caption{An illustration on comparison of tail probabilities in a Gaussian Q-function.}
	\label{fig:Qfun_plot_tail_probty_comp}
\end{figure}

{\em Remarks:} We see that player$-1$'s skill level is higher if
\begin{equation}
	\label{eq:kPt_score_prbty_P2vsP1}
	Q\brac{\frac{k - S_{1}p_{1}}{\sqrt{S_{1}p_{1} (1-p_{1})}}} > Q\brac{\frac{k - S_{2}p_{2}}{\sqrt{S_{2}p_{2} (1-p_{2})}}}.
\end{equation}
Let $y_1 > 0$ and $y_2 > 0$. Note that for $y_1 < y_2$, $Q(y_1) > Q(y_2)$, which is illustrated in figure~\ref{fig:Qfun_plot_tail_probty_comp}.

Therefore,  we see that player$-1$'s skill level is higher if
\begin{equation}
	\label{eq:P2vsP1_simple_rule}
	\frac{k - S_{1}p_{1}}{\sqrt{S_{1}p_{1} (1-p_{1})}} < \frac{k - S_{2}p_{2}}{\sqrt{S_{2}p_{2} (1-p_{2})}}.
\end{equation}

Suppose $p_{1} \approx p_{2} = p$. We find that the condition simplifies to 
\begin{equation}
	\label{eq:P2eqP1_simple_rule}
	\brac{\frac{k - S_{1}p}{k - S_{2}p}}^{2} < \frac{S_1}{S_{2}}.
\end{equation}

\subsection{Modeling and Skill Comparison of Two Teams}
\label{sec:prob_anal_team}

We now extend the modeling and analysis to two teams. Consider the squash game being played by two teams. Let the two teams be denoted by team$-$A and team$-$B. Let $S_{A1}$, $S_{A2}$ denote the number of squash shots played by players one and two of team$-$A, respectively. Similarly, Let $S_{B1}$,  $S_{B2}$ denote the number of squash shots played by players one and two of team$-$B, respectively. To analyze the two team squash game, We follow a similar modeling strategy mentioned in the previous section. 

{\em Probabilistic modeling ans skill comparison rule:} Let $W_{A1} \define \sum_{s_{a1} = 0}^{S_{A1}} I_{s_{a1}}$. Let $W_{A2} \define \sum_{s_{a2} = 0}^{S_{A2}} I_{s_{a2}}$. By using the Gaussian approximation, we see that the distribution of $W_{A1}$ is as follows. $W_{A1} \sim \mathcal{N}\brac{S_{a1}p_{a1}, S_{a1}p_{a1} (1-p_{a1})}$. Further, $W_{A2} \sim \mathcal{N}\brac{S_{a2}p_{a2}, S_{a2}p_{a2} (1-p_{a2})}$. Let $W_{A} = W_{A1} + W_{A2}$. Note that $W_{A1}$ and $W_{A2}$ are statistically independent. Clearly, $W_{A}$ is also Gaussian distributed, that is, $W_{A} \sim \mathcal{N}\brac{S_{a1}p_{a1} + S_{a2}p_{a2}, S_{a1}p_{a1} (1-p_{a1}) + S_{a2}p_{a2} (1-p_{a2})}$.

We are interested in determining the probability of scoring $k$ points by the team. Mathematically, we write $\mathrm{P}(W_{A} > k)$. Assuming Gaussian approximation~\cite{Papoulis_book} for a large number of squash shots, we have 
\begin{equation}
	\label{eq:kPt_score_prbty_teamA}
	\mathrm{P}(W_{A} > k) \approx Q\brac{\frac{k - \brac{S_{a1}p_{a1} + S_{a2}p_{a2}}}{\sqrt{S_{a1}p_{a1} (1-p_{a1}) + S_{a2}p_{a2} (1-p_{a2})}}},
\end{equation}

Similarly, for Team$-$B, we can show that
\begin{equation}
	\label{eq:kPt_score_prbty_teamB}
	\mathrm{P}(W_{B} > k) \approx Q\brac{\frac{k - \brac{S_{b1}p_{b1} + S_{b2}p_{b2}}}{\sqrt{S_{b1}p_{b1} (1-p_{b1}) + S_{b2}p_{b2} (1-p_{b2})}}}.
\end{equation}

{\em Remarks:} We see that Team$-$A's skill level is higher if
\begin{multline}
	\label{eq:kPt_score_prbty_TeamA_vs_TeamB}
	Q\brac{\frac{k - \brac{S_{a1}p_{a1} + S_{a2}p_{a2}}}{\sqrt{S_{a1}p_{a1} (1-p_{a1}) + S_{a2}p_{a2} (1-p_{a2})}}} \\ > Q\brac{\frac{k - \brac{S_{b1}p_{b1} + S_{b2}p_{b2}}}{\sqrt{S_{b1}p_{b1} (1-p_{b1}) + S_{b2}p_{b2} (1-p_{b2})}}}.
\end{multline}

Note that for $z_1 < z_2$, $Q(z_1) > Q(z_2)$. 
Therefore,  we see that team$-$A's skill level is higher if
\begin{equation}
	\label{eq:TeamA_vs_TeamB_simple_rule}
	\frac{k - \brac{S_{a1}p_{a1} + S_{a2}p_{a2}}}{\sqrt{S_{a1}p_{a1} (1-p_{a1}) + S_{a2}p_{a2} (1-p_{a2})}} < \frac{k - \brac{S_{b1}p_{b1} + S_{b2}p_{b2}}}{\sqrt{S_{b1}p_{b1} (1-p_{b1}) + S_{b2}p_{b2} (1-p_{b2})}}.
\end{equation}

Alternatively, the skill comparison rule is given by
\begin{equation}
	\label{eq:TeamA_vs_TeamB_alt_rule}
	\frac{k - \brac{S_{a1}p_{a1} + S_{a2}p_{a2}}}{k - \brac{S_{b1}p_{b1} + S_{b2}p_{b2}}} < \sqrt{\frac{{S_{a1}p_{a1} (1-p_{a1}) + S_{a2}p_{a2} (1-p_{a2})}}{{S_{b1}p_{b1} (1-p_{b1}) + S_{b2}p_{b2} (1-p_{b2})}}}.
\end{equation}

\subsection{Numerical evaluation and Performance plot}
\label{sec:numEval_plot}

Consider the two player squash game. We numerically evaluate the $k-$point scoring probability for the following parameters. We set $p = 0.18$, $S = 40$. We plot the probability as a function of score points $k$.

\begin{figure}[htb!]
	\centering \includegraphics[width=1\linewidth]{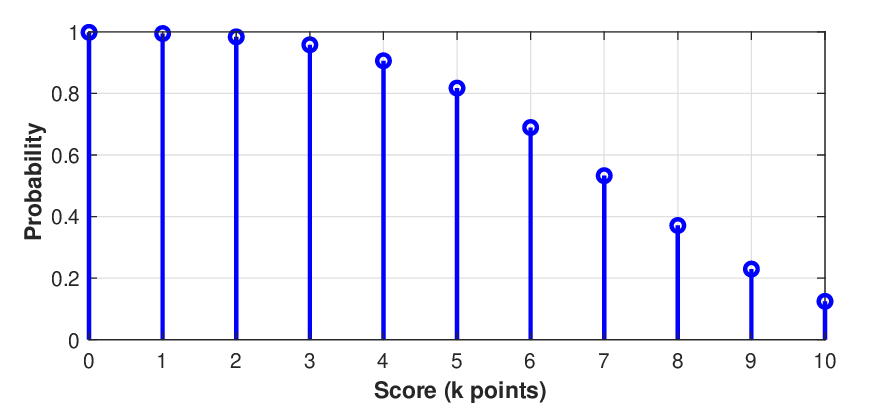}
	\caption{Points scoring probability as a function of $k$.}
	\label{fig:Prbty_vs_k_stem}
\end{figure}

Figure~\ref{fig:Prbty_vs_k_stem} plots scoring probability as a function of $k$. We see that the probability decreases as $k$ increases. This decreasing trend in probability is intuitively satisfying because each player's understanding of the opponent's strategy and strength will improve as the game evolves. Thus, for the equally skillful players, this will lead to more squash shots to get more points before the game ends.

\section{Analysis of Shot Types: A Case Study}
\label{sec:sim_res}

\begin{figure}[h]
	\centering
	\includegraphics[width=0.9\textwidth]{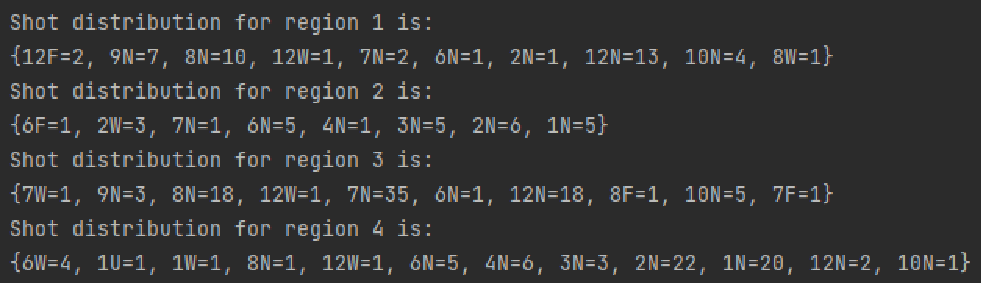}
	\caption{Output of the Java-based shot-analysis tool for a single professional match.}
	\label{java_snip}
\end{figure}

A Java program was developed to analyze shot distributions across the four regions
$R1$, $R2$, $R3$, and $R4$ of the squash court.  
Figure~\ref{java_snip} shows a sample analysis for one professional match.  
The first number (1--12) indicates the shot type, followed by a letter denoting the shot
outcome. Each combination is accompanied by its corresponding count.  
The outcomes are stored region-wise, allowing for a detailed breakdown of shot tendencies.
The complete dataset and code repository are available at~\cite{github}.

We now present the region-wise shot-distribution (SD) results for professional players.

\subsection{Region-wise Shot Distribution (SD) for Professional Players}
\label{subsec:Shot_distbn}

Tables~\ref{table:1} and~\ref{table:2} summarize the shot types executed by professional
players across the four regions, along with their success and failure counts.
A total of $900$ shots were analyzed from four professional matches.

\begin{table}[htb!]
	\caption{Shot types in Region~1 and Region~2 (professional players).}
	\centering
	\begin{tabular}{||c||c c c||}
		\hline
		Region & Shot Type & Successful Shots & Unsuccessful Shots \\ [0.5ex]
		\hline\hline
		$R1$ & Backhand Parallel & $14$ & $0$ \\
		$R1$ & Backhand Cross & $32$ & $1$ \\
		$R1$ & Backhand Drop & $41$ & $6$ \\
		$R1$ & Backhand Lob & $22$ & $1$ \\
		$R1$ & Other & $3$ & $1$ \\ [1ex]
		\hline\hline
		$R2$ & Forehand Parallel & $9$ & $0$ \\
		$R2$ & Forehand Cross & $30$ & $1$ \\
		$R2$ & Forehand Drop & $13$ & $2$ \\
		$R2$ & Forehand Lob & $8$ & $0$ \\
		$R2$ & Other & $6$ & $0$ \\ [1ex]
		\hline
	\end{tabular}
	\label{table:1}
\end{table}

\begin{table}[htb!]
	\caption{Shot types in Region~3 and Region~4 (professional players).}
	\centering
	\begin{tabular}{||c||c c c||}
		\hline
		Region & Shot Type & Successful Shots & Unsuccessful Shots \\ [0.5ex]
		\hline\hline
		$R3$ & Backhand Parallel & $323$ & $2$ \\
		$R3$ & Backhand Cross & $80$ & $1$ \\
		$R3$ & Backhand Drop & $77$ & $4$ \\
		$R3$ & Backhand Boast & $19$ & $1$ \\
		$R3$ & Other & $12$ & $0$ \\ [1ex]
		\hline\hline
		$R4$ & Forehand Parallel & $97$ & $2$ \\
		$R4$ & Forehand Cross & $98$ & $1$ \\
		$R4$ & Forehand Drop & $26$ & $1$ \\
		$R4$ & Forehand Boast & $19$ & $2$ \\
		$R4$ & Other & $17$ & $1$ \\ [1ex]
		\hline
	\end{tabular}
	\label{table:2}
\end{table}

\subsection{Region-wise SD for Intermediate-Level Players}

Tables~\ref{table:3} and~\ref{table:4} present the corresponding analysis for
intermediate-level (IL) players.  
A total of $500$ shots were analyzed across four IL matches.

\begin{table}[htb!]
	\caption{Shot types in Region~1 and Region~2 (intermediate-level players).}
	\centering
	\begin{tabular}{||c||c c c||}
		\hline
		Region & Shot Type & Successful Shots & Unsuccessful Shots \\ [0.5ex]
		\hline\hline
		$R1$ & Backhand Parallel & $5$ & $1$ \\
		$R1$ & Backhand Cross & $10$ & $2$ \\
		$R1$ & Backhand Drop & $12$ & $5$ \\
		$R1$ & Backhand Lob & $3$ & $1$ \\
		$R1$ & Other & $5$ & $1$ \\ [1ex]
		\hline\hline
		$R2$ & Forehand Parallel & $14$ & $3$ \\
		$R2$ & Forehand Cross & $11$ & $1$ \\
		$R2$ & Forehand Drop & $12$ & $2$ \\
		$R2$ & Forehand Lob & $12$ & $2$ \\
		$R2$ & Other & $4$ & $0$ \\ [1ex]
		\hline
	\end{tabular}
	\label{table:3}
\end{table}

\begin{table}[htb!]
	\caption{Shot types in Region~3 and Region~4 (intermediate-level players).}
	\centering
	\begin{tabular}{||c||c c c||}
		\hline
		Region & Shot Type & Successful Shots & Unsuccessful Shots \\ [0.5ex]
		\hline\hline
		$R3$ & Backhand Parallel & $108$ & $11$ \\
		$R3$ & Backhand Cross & $52$ & $2$ \\
		$R3$ & Backhand Drop & $23$ & $4$ \\
		$R3$ & Backhand Boast & $9$ & $1$ \\
		$R3$ & Other & $6$ & $2$ \\ [1ex]
		\hline\hline
		$R4$ & Forehand Parallel & $71$ & $4$ \\
		$R4$ & Forehand Cross & $42$ & $1$ \\
		$R4$ & Forehand Drop & $8$ & $1$ \\
		$R4$ & Forehand Boast & $18$ & $2$ \\
		$R4$ & Other & $7$ & $1$ \\ [1ex]
		\hline
	\end{tabular}
	\label{table:4}
\end{table}

\begin{figure}[htb!]
	\centering
	\begin{subfigure}{0.45\columnwidth}
		\centering
		\begin{tikzpicture}
			\tikzstyle{every node}=[font=\small]
			\pie [rotate=180, radius=1.8]{61/W, 18/FE, 21/UFE}
		\end{tikzpicture}
		\caption{Professional-level match: Winner (61\%), Forced Error (18\%), Unforced Error (21\%).}
		\label{fig:shot_ooutcome_proflevel}
	\end{subfigure}
	\quad
	\begin{subfigure}{0.45\columnwidth}
		\centering
		\begin{tikzpicture}
			\tikzstyle{every node}=[font=\small]
			\pie [rotate=180, radius=1.8]{46/W, 18/FE,36/UFE}
		\end{tikzpicture}
		\caption{Intermediate-level match: Winner (46\%), Forced Error (18\%), Unforced Error (36\%).}
		\label{fig:shot_ooutcome_intlevel}
	\end{subfigure}
	\caption{Shot-outcome distributions for professional and intermediate players.}
	\label{fig:TOF}
\end{figure}
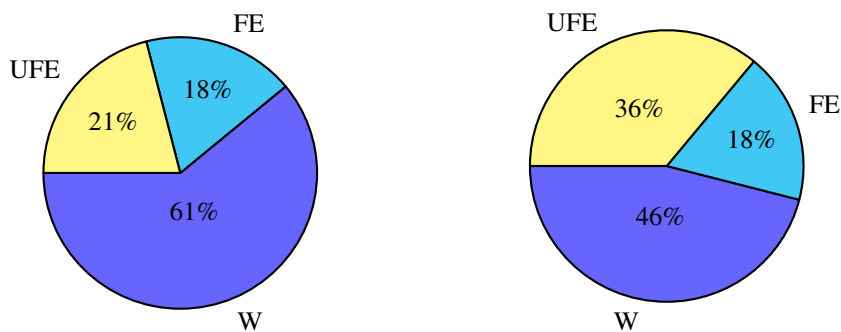

\subsection{Shot Outcome for Professional-Level Matches}

The overall shot-outcome distribution for professional players is shown in
Figure~\ref{fig:shot_ooutcome_proflevel}.  
The chart highlights the proportion of winners, forced errors, and unforced errors.

\subsection{Shot Outcome for Intermediate-Level Matches}

Similarly, the shot-outcome distribution for intermediate-level players is illustrated in
Figure~\ref{fig:shot_ooutcome_intlevel}.  
A clear difference in outcome tendencies is visible when compared with professional-level performance.

\section{Analysis of Shot Output}
\label{sec:Analysis_output}

We now analyze the outcomes of three key shot categories: winner shots, forced-error shots, and unforced-error shots. The results highlight clear differences in shot selection, control, and execution between professional-level (PL) and intermediate-level (IL) players.

\subsection{Winner Shot Analysis}

At the PL, most winners were generated from the BH drop shot (40\%), followed by the FH drop shot (24\%). In contrast, at the IL, the majority of winners arose from the FH parallel drive (58\%). This distinction underscores the greater technical proficiency of professional players: executing a drop shot, especially from the backhand side, demands substantially higher precision and control than producing a forehand parallel drive.

\subsection{Forced-Error Shot Analysis}

Most forced errors at the PL originated from the BH drop shot (42\%), followed by the BH boast shot (17\%). At the IL, the highest proportion of forced errors came from the BH parallel drive (34\%). These trends suggest that professionals pressure opponents by employing technically demanding shots, whereas intermediates induce errors primarily through basic directional drives.

\subsection{Unforced-Error Shot Analysis}

At the PL, the highest percentage of unforced errors also came from the BH drop shot (53\%), followed by the FH boast shot (15\%). Similarly, at the IL, the largest share of unforced errors stemmed from the BH drop shot (40\%). This consistency across levels reflects the inherent difficulty of the backhand drop: even experienced players face challenges in maintaining accuracy and depth on this shot.

\section{Concluding Remarks}
\label{sec:conclusions}

We proposed a simple yet effective probabilistic model for both two-player and two-team squash games. A probability-based skill comparison rule was developed to assess individual and team performance by analyzing the probabilities associated with point scoring. In parallel, we examined the distribution of squash shots across different court regions, comparing patterns between professional-level (PL) and intermediate-level (IL) players using data collected on shot type and shot location.

Our data-driven analysis revealed that 61\% of points at the PL resulted from winners, compared to 46\% at the IL, which aligns with the expectation that higher-skilled players finish points more efficiently. A clear distinction between skill levels was also observed in unforced errors: PL players recorded 21\%, whereas IL players exhibited 36\%. Region-wise and shot-wise evaluations showed that players at both levels operated predominantly in the backcourt, underscoring the strategic importance of the backhand parallel drive. The backhand drop shot emerged as a high-risk, high-reward move responsible for a large fraction of professional winners but also a primary source of unforced errors across both groups. 

Overall, the probabilistic modeling and empirical observations presented in this work offer useful insights into squash dynamics and player behavior. These findings could support and accelerate further developments in sports analytics, performance evaluation, and data-driven coaching.

\section*{Declarations}

\textbf{Funding:} No funding was received for this work.\\
\textbf{Competing interests:} The authors declare no competing interests.\\
\textbf{Authors' contributions:} All authors contributed to the work.\\
\textbf{Availability of data and materials:} The datasets and code are available at \url{https://github.com/PrathameshAnwekar/SOP-SquashProbabilisticAnalysis}.

\end{document}